\documentclass[aps,prl,twocolumn,showpacs]{revtex4}
\usepackage{graphicx}

\begin{document}

\title{Spontaneous formation of a \boldmath$\pi$ soliton in a superconducting
  wire with an odd number of electrons}

\author{Hyok-Jon Kwon and Victor M. Yakovenko}

\affiliation{Department of Physics and Center for Superconductivity
Research, University of Maryland, College Park, Maryland 20742-4111}

\date{{\bf cond-mat/0204033}, v.1 1 April 2002, v.2 7 June 2002}

\begin{abstract}

We consider a one-dimensional superconducting wire where the total
number of electrons can be controlled in the Coulomb blockade regime.
We predict that a $\pi$ soliton (kink) will spontaneously form in the
system when the number of electron is odd, because this configuration
has a lower energy.  If the wire with an odd number of electrons is
closed in a ring, the phase difference on the two sides of the soliton
will generate a supercurrent detectable by SQUID.  The two degenerate
states with the current flowing clockwise or counterclockwise can be
utilized as a qubit.

\end{abstract} 


\maketitle

\paragraph{Introduction.}

In a mesoscopic sample in the Coulomb blockage regime, the total
number of electrons can be controlled discretely by gate voltage
\cite{blockade}.  It was predicted theoretically \cite{Averin93} and
verified experimentally \cite{Tinkham92} that the ground-state
energies of a superconducting sample with even and odd number of
electrons differ by the superconducting gap $\Delta_0$, because the
odd electron cannot form a Cooper pair.  In the present paper, we
discuss another manifestation of the even-odd asymmetry in a
one-dimensional (1D) superconducting wire or constriction.  We show
that when such a system has an odd number of electrons, it can lower
its energy by spontaneously forming a kink soliton of the order
parameter, analogously to the charge-density-wave solitons in
polyacetylene \cite{Brazovskii,Takayama,Schrieffer} and other systems
with similar broken symmetry \cite{Brazovskii-Moriond}.  If the wire
is closed in a ring, the phase difference on the two sides of the
soliton would produce a supercurrent in the ring, which could be
detected by a sensitive SQUID.  The two degenerate states with the
current flowing clockwise or counterclockwise can be considered as a
qubit \cite{qubit}.

\paragraph{1D superconductor with a kink soliton.}

Let us consider a 1D $s$-wave BCS superconductor characterized by the
mean-field Hamiltonian density
\begin{eqnarray}
   && \hat H(x) = 
   \hat c^\dag_\uparrow(x)\,\hat\xi\,\hat c_\uparrow(x)
   + \hat c^\dag_\downarrow(x)\,\hat\xi\,\hat c_\downarrow(x)
\label{eq:MFT} \\
   && {}+ \hat c_\downarrow(x)\,\Delta^*(x)\,\hat c_\uparrow(x)
   + \hat c_\uparrow^\dag(x)\,\Delta(x)\,
   \hat c_\downarrow^\dag(x) - |\Delta (x)|^2/g.
\nonumber
\end{eqnarray}
Here $\hat c_{\uparrow,\downarrow}^{(\dag)}(x)$ are the creation and
destruction operators of electrons with the spin projections
$\uparrow$ and $\downarrow$ at the point $x$, $\Delta(x)$ is the
superconducting pairing potential satisfying the self-consistency
condition $\Delta(x)=g\langle \hat c_{\downarrow}(x)\hat
c_{\uparrow}(x)\rangle$, $g<0$ is the BCS coupling constant, and
$\hat{\xi}$ is the kinetic energy operator.  Near the Fermi energy,
the electron dispersion relation can be linearized:
$\hat\xi\approx-\alpha iv_F\partial_x$, where $\alpha=\pm$ labels the
right- and left-moving electrons with the momenta near the two Fermi
points $\alpha k_F$, $v_F$ is the Fermi velocity, and the Planck
constant is $\hbar=1$.  All calculations in the paper are done at zero
temperature.

The eigenstates of Hamiltonian (\ref{eq:MFT}) are the Bogoliubov
quasiparticles characterized by the wave functions
$\psi_n^{(\alpha)}(x)=[u_n^{(\alpha)}(x),v_n^{(\alpha)}(x)]$ that
satisfy the Bogoliubov-de Gennes (BdG) equation \cite{books}
\begin{equation}
   \left(\begin{array}{cc} 
   -\alpha iv_F \partial_x & \Delta(x) \\ 
   \Delta^*(x) &\alpha iv_F \partial_x
   \end{array}\right) 
   \psi_n^{(\alpha)}(x) = E_n^{(\alpha)} \psi_n^{(\alpha)}(x)
\label{eq:BdG}
\end{equation}
with the eigenenergies $E_n^{(\alpha)}$.  The spectrum of Eq.\
(\ref{eq:BdG}) consists of solutions with positive and negative
energies related by charge conjugation.  The superconducting
ground-state energy $F$ relative to the normal state energy and the
BCS self-consistency relation can be expressed in terms of these
solutions \cite{books}:
\begin{eqnarray}
   & F =\sum_{n,\alpha} 
   \left( E_n^{(\alpha)} + |\xi_n^{(\alpha)}| \right)
   -\int dx\,|\Delta(x)|^2/g, &
\label{eq:UG} \\
   & \Delta(x)=-g\sum_{n,\alpha} 
   u_n^{(\alpha)}(x)\,v_n^{(\alpha)*}(x), &
\label{eq:gap}
\end{eqnarray}
where the sums are taken over the occupied states with
$E_n^{(\alpha)}<0$.

\begin{figure}
\includegraphics[width=0.8\linewidth,angle=0]{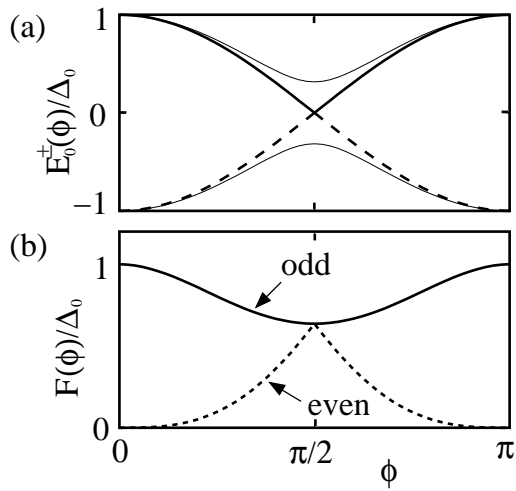} 
\caption{ (a) The energies $E_0^{(\alpha)}$ of the subgap bound states
  vs phase $\phi$.  The thick solid and dashed lines correspond to the
  transmission coefficient $\tau=1$, and the thin lines to $\tau=0.9$.
  (b) The total energy $F$ of the one-channel superconducting wire
  with an odd (solid line) and an even (dashed line) number of
  electrons vs phase $\phi$, calculated for $\tau=1$.}
\label{fig:JJ} 
\end{figure}

Suppose the pairing potential has a phase soliton (kink) at $x=0$, so
that $\Delta(x)\to\Delta_0 e^{\pm i\phi}$ at $x\to\pm\infty$:
\begin{eqnarray}
  & \Delta(x)=\Delta_1 + i\Delta_2(x), \quad
  \Delta_1 = \Delta_0 \cos\phi, &
\label{eq:Delta} \\
  & \Delta_2(x)=\kappa v_F \tanh \kappa x, \quad
  \kappa = \Delta_0\sin\phi/v_F. &
\label{eq:kappa}
\end{eqnarray}
In this case \cite{Furusaki,Feigel'man}, the spectrum of Eq.\
(\ref{eq:BdG}) has a continuous part $\pm E_q^{(\alpha)}$, where $q$
is the quasiparticle momentum counted from $\alpha k_F$, and discrete
subgap states $E_0^{(\alpha)}$:
\begin{equation}
   E_q^{(\alpha)}=\sqrt{(v_Fq)^2+\Delta_0^2}, \quad
   E_0^{(\alpha)}=-\alpha\Delta_0\cos\phi.
\label{eq:E_0}  
\end{equation}
The energies $E_0^\pm(\phi)$ of the subgap states (\ref{eq:E_0}) are
shown in Fig.\ \ref{fig:JJ}(a) by the thick solid and dashed lines.
The total energy $F$, given by Eq.\ (\ref{eq:UG}), can be separated
into the contributions from the continuous spectrum and from the
discrete states.  The former contribution, counted from the ground
state energy of a uniform BCS superconductor with an even number of
electrons, has the form
\begin{equation}
   F_c=\Delta_0 - 2\sum_{q_j}\left(E_{q_j}-E_{\bar q_j}\right)
   -\int dx\,\frac{|\Delta(x)|^2-\Delta_0^2}{g}.
\label{eq:UB}
\end{equation}
The first term in the r.h.s.\ of Eq.\ (\ref{eq:UB}) reflects the fact
that one state has been removed from the continuum to the discrete
spectrum.  The factor 2 in front of the second term represents the
equal contributions from the states with $\alpha=\pm$.  In the second
term, $q_j=\bar q_j-\zeta_j/L$ is the quasiparticle momentum displaced
by the phase shift $\zeta_j$ from the unperturbed value $\bar q_j=2\pi
j/L$ for a uniform system without a kink.  (Here the periodic boundary
condition over the length $L$ is implied, and $j$ is an integer.)  The
calculations of the phase shifts and the dependence of $F_c$ on $\phi$
are given at the end of the paper (see also
\cite{Brazovskii,Takayama,Neveu}).  The result is
\begin{equation}
   F_c(\phi)={2\Delta_0 \over \pi} \left[ 
   \left({\pi\over2} -\phi\right)\cos\phi +\sin\phi \right].
\label{eq:F_c}
\end{equation}

Let us examine how the ground state energies $F_e(\phi)$ and
$F_o(\phi)$ of the system with an even and an odd number of electrons
depend on the phase difference $2\phi$.  At $\phi=0$, the spectrum has
only the continuous part, which is completely filled below $-\Delta_0$
and completely empty above $+\Delta_0$ when the number of electrons is
even.  As $\phi$ increases, the discrete state (\ref{eq:E_0}) with the
negative energy remains occupied, whereas the one with the positive
energy is empty.  Thus,
\begin{equation}
   F_e(\phi)=F_c(\phi)-\Delta_0|\cos\phi|.
\label{eq:F_e}  
\end{equation}
As shown in Fig.\ \ref{fig:JJ}(b) by the dashed line, $F_e(\phi)$ is
minimal at $\phi=0$.  Thus, the system with an even number of
electrons prefers a uniform state.

On the other hand, when the number of electrons is odd, the
quasiparticle state with the energy $+\Delta_0$ is filled by the odd
electron at $\phi=0$ \cite{Averin93}.  Thus, as $\phi$ increases, both
the lower and upper discrete states (\ref{eq:E_0}) stay occupied.
Because their energies cancel, the total energy in the odd case is
\begin{equation}
  F_o(\phi)=F_c(\phi),
\label{eq:U_o}
\end{equation}
which is shown by the solid line in Fig.\ \ref{fig:JJ}(b).  In
contrast to the even case, $F_o(\phi)$ decreases with increasing
$\phi$ and has a minimum at $\phi=\pi/2$ with the energy
$E_s=F_o(\pi/2)=2\Delta_0/\pi<\Delta_0$, lower than in the uniform
state.  Thus, a 1D superconductor with an odd number of electrons
should spontaneously create a kink with the phase difference
$2\phi=\pi$ in order to decrease the total energy.  This phenomenon is
analogous to creation of a soliton in a charge- or spin-density wave
with an odd number of electrons
\cite{Brazovskii,Takayama,Schrieffer,Brazovskii-Moriond}.

Now let us consider the case where the 1D superconducting wire has $N$
channels, e.g.\ consists of $N$ parallel chains with the common
potential $\Delta(x)$.  When the number of electrons is odd, $N$
discrete states with negative energies (\ref{eq:E_0}) and one with
positive energy are filled.  So, the total energy of the system is
$F_o(\phi)=NF_c(\phi)-(N-1)\Delta_0|\cos\phi|$, and the minimum
$(\Delta_02N/\pi)\sin(\pi/2N)$ is achieved at $2\phi=\pi/N$.  Thus,
the soliton also forms in this case, but with the phase difference
reduced by $1/N$.  When $N\gg1$, the energy gain is insignificant, and
soliton formation becomes irrelevant.

\paragraph{Superconducting ring with a kink.}
Now suppose that the 1D superconducting wire (with $N=1$) is closed in
a ring of length $L$.  The length $L$ is assumed to be much longer
than the coherence length $\xi=\hbar v_F/\Delta_0$, so the soliton
energy $E_s$ is not significantly affected by closing of the wire.  On
the other hand, $L$ should be short enough, so that the whole ring is
in the Coulomb blockade regime, and an external gate can control the
total number of electrons to be odd.  The latter condition requires
that $\Delta_0<E_c$ \cite{Averin93}, where $E_c\sim e^2/D$ is the
Coulomb charging energy, and $D$ is the overall size of the sample.
For a simple ring, $D=L$, but in general $D\agt L$, as discussed
below.  Thus, we need
\begin{equation}
   \xi=\hbar v_F/\Delta_0 \ll L \alt D \alt e^2/\Delta_0,
\label{eq:L}  
\end{equation}
so the Fermi velocity has to be much lower than the Bohr velocity:
$v_F\ll e^2/\hbar=2.2\times10^8$ cm/s.  This condition is satisfied in
a typical metal with $v_F\sim 10^7$ cm/s.

When the number of electrons is odd, the soliton creates the phase
difference $2\phi=\pi$ between the ends of an open wire.  However,
when the wire is closed, the superconducting phase $\theta$ must
continuously interpolate between the phases $\pm\pi/2$ at the opposite
sides of the soliton, so $\Delta(x)=\Delta_0e^{i\theta(x)}$.  The
small phase gradient $|\partial\theta/\partial x|=\pi/L$ creates the
supercurrent
\begin{equation}
   J=\pm\frac{en_s}{2m}\,\frac{\partial\theta}{\partial x}
   =\pm\frac{ev_F}{L},
\label{eq:J}  
\end{equation}
where $n_s=4k_F/2\pi$ is the superfluid density equal to the total
electron density at zero temperature \cite{current}.  Notice that the
supercurrent $J$ can flow in the ring either clockwise or
counterclockwise, thus the time-reversal symmetry is spontaneously
broken.  These two degenerate states can be viewed as a qubit
\cite{qubit}.  The supercurrent (\ref{eq:J}) is equal to the maximal
persistent current in a normal (nonsuperconducting) ring with an odd
number of electrons \cite{Gefen}.  It corresponds to the odd electron
of charge $e$ moving with the Fermi velocity $v_F$ around the ring of
circumference $L$.  This current creates the magnetic flux $\Phi \sim
JL/c \sim ev_F/c \sim 10^{-5}\Phi_0$ (where $\Phi_0=\pi\hbar c/e$ is
the flux quantum), which could be detected, in principle, by a
sensitive SQUID.  \emph{We predict that, when the number of electrons
in the ring is switched between odd and even by an external gate, the
current (\ref{eq:J}) should appear and disappear in the ring.}  In the
case of $N$ channels, the current is given by the same Eq.\
(\ref{eq:J}).

Above we considered an idealized situation of a uniform
superconducting ring.  However, the conclusion also applies to other,
more realistic experimental setups.  Suppose a straight 1D
superconducting wire of length $L$, e.g.\ a carbon nanotube
\cite{nanotube}, is shunted by a thick conventional superconductor of
size $D$.  In this case, the soliton and the phase gradient are
localized within the 1D wire, whereas the phase gradient over the
shunt is close to zero, so the supercurrent is given by Eq.\
(\ref{eq:J}).  Another setup is a narrow constriction with a small
number $N$ of conducting channels in a thick superconducting ring.  In
this case, $L$ is the length of the 1D constriction, and $D$ is the
overall size of the ring.  In general, when the ring is nonuniform,
the transmission coefficient $\tau$ through the weak link is less than
one.  In this case, the energies of the subgap states are
$E_0=\pm\Delta_0\sqrt{1-\tau\sin^2\phi}$ \cite{Furusaki,Feigel'man},
shown by the thin lines in Fig.\ \ref{fig:JJ}(a).  Although we cannot
calculate $F_c(\phi)$ exactly when $\tau\neq1$, the total energy of
the system should be close to Eq.\ (\ref{eq:F_c}) if $\tau\approx1$,
and the kink soliton should be still energetically favorable.  Thus,
our theory should be applicable to the atomic contacts with $N\sim1$
and $\tau\approx1$ realized in experiments \cite{Urbina}.

\paragraph{Calculation of $F_c(\phi)$ via phase shifts.}

To simplify presentation, we write the equations below specifically
for $\alpha=+$.  In order to calculate the phase shifts, let us
introduce the wave functions $f_\pm(x)=u(x)\pm v(x)$. (Notice that
this index $\pm$ is different from $\alpha$.)  Then the BdG equation
(\ref{eq:BdG}) acquires the supersymmetric form \cite{SUSY,UIUC}
\begin{eqnarray}
  (E\mp\Delta_1) f_\pm=-iv_F\,\partial f_\mp/\partial x
   \mp i\Delta_2(x) f_\mp, && 
\label{eq:BdGpm} \\
   \left(E^2+v_F^2\,\frac{\partial^2}{\partial x^2}-|\Delta(x)|^2\mp 
   v_F\,\frac{\partial\Delta_2(x)}{\partial x}\right)f_\pm=0.&& 
\label{eq:BdGdiagonal}
\end{eqnarray}
Substituting Eqs.\ (\ref{eq:Delta}) and (\ref{eq:kappa}) into Eqs.\ 
(\ref{eq:BdGpm}) and (\ref{eq:BdGdiagonal}), we find the
(unnormalized) wave functions
\begin{equation}
   f_+(x)=e^{iqx}, \quad 
   f_-(x)=\frac{v_Fq+i\Delta_2(x)}{E+\Delta_1}e^{iqx}.
\label{eq:f}  
\end{equation}
From Eq.\ (\ref{eq:f}), we observe that, when $x$ changes from
$-\infty$ to $+\infty$, $f_-$ acquires the phase factor
$\Xi=(q+i\kappa)/(q-i\kappa)$.  In order to impose periodic boundary
conditions consistently, let us assume that the system also has an
antikink far away from $x=0$.  In a similar manner, it is easy to show
that $f_+$ acquires the same phase factor $\Xi$ upon traversing the
antikink.  Thus, when $x$ changes from one boundary of the sample to
another, both $f_\pm$ experience the phase shift
$\zeta_q=\arctan(\kappa/q)$ per \emph{one} soliton
\cite{Takayama,Neveu}.  The same result can be also obtained by
imposing zero boundary conditions at the ends of the wire.

Using the relation $q_j=\bar q_j-\zeta_j/L$, we write Eq.\
(\ref{eq:UB}) as
\begin{equation}
   F_c=\Delta_0 +
   2\int\frac{dq}{2\pi}\,\zeta_q\,\frac{\partial E_q}{\partial q}
   +\frac{2\kappa v_F^2}{g},
\label{eq:shift}  
\end{equation}
where we took the integral over $x$ using Eqs.\ (\ref{eq:Delta}) and
(\ref{eq:kappa}).  Integrating by parts over $q$ and taking into
account the BCS self-consistency relation
\begin{equation}
   -\frac1g=\int \frac{dq}{2\pi E_q},
\label{BCS}
\end{equation}
we rewrite Eq.\ (\ref{eq:shift}) as follows
\begin{equation}
   F_c = \frac{2v_F\kappa}{\pi} - 2\int\frac{dq}{2\pi}\left(
   E_q\,\frac{\partial\zeta_q}{\partial q}
   +\frac{v_F^2\kappa}{E_q}\right).
\label{eq:Fc-zeta}  
\end{equation}
Substituting $\zeta_q=\arctan(\kappa/q)$ into Eq.\ (\ref{eq:Fc-zeta})
and taking the integral over $q$, as in Ref.\ \cite{Neveu}, we obtain
Eq.\ (\ref{eq:F_c}).

\paragraph{An alternative calculation of $F_c(\phi)$.}

Following the method of Ref.\ \cite{Brazovskii}, let us calculate the
derivative
\begin{equation}
   \frac{dF_c(\phi)}{d\phi} = \int dx 
   \left[ {\delta F_c \over \delta\Delta^*(x)}
   {\partial\Delta^*(x) \over \partial\phi} + {\rm c.c.}
   \right].
\label{eq:dU/dphi}
\end{equation}
We write the eigenenergies of the BdG equation (\ref{eq:BdG}) as
\begin{eqnarray}
   E_n^{(\alpha)} = \int dx\,
   [u_n^{(\alpha)*}(x)\,\hat\xi\,u_n^{(\alpha)}(x)
   -v_n^{(\alpha)}(x)\,\hat\xi\,v_n^{(\alpha)*}(x) && 
\nonumber \\
   {}+u_n^{(\alpha)}(x)\,\Delta^*(x)\,v_n^{(\alpha)*}(x)  
   + v_n^{(\alpha)}(x)\,\Delta(x)\,u_n^{(\alpha)*}(x)], && 
\label{eq:En}
\end{eqnarray}
where the normalization condition $\int dx(|u|^2+|v|^2)=1$ is implied.
Taking variational derivatives of Eqs.\ (\ref{eq:UB}) and
(\ref{eq:En}) with respect to $\Delta^*(x)$, we find
\begin{equation}
   {\delta F_c \over \delta\Delta^*(x)} = 
   -\sum_{q,\alpha} u_q^{(\alpha)}(x)\, v_q^{(\alpha)*}(x)
   -{\Delta(x) \over g}.
\label{eq:FUG}
\end{equation}
The normalized wave functions
$\psi_q^{(\alpha)}=(u_q^{(\alpha)},v_q^{(\alpha)})$ can be deduced
from Eq.\ (\ref{eq:f})
\begin{equation}
   \psi_q^{(\alpha)} =  
   \sqrt{E_q+\alpha\Delta_1 \over 4LE_q}
   \left( \begin{array}{c} 
   1+ \alpha {v_Fq+i\Delta_2(x) \over E_q + \alpha\Delta_1} \\ 
   \alpha - {v_Fq+i\Delta_2(x) \over E_q + \alpha\Delta_1}
   \end{array} \right) e^{iqx}.
\label{eq:cont}
\end{equation}
Using Eq.\ (\ref{eq:cont}), we find
\begin{equation}
   \sum_{q,\alpha} u_q^{(\alpha)} v_q^{(\alpha)*} 
   = \Delta(x) \int{dq \over 2\pi E_q} 
   - {\kappa(\pi/2 -\phi) \over 2\pi\cosh^2(\kappa x)}.
\label{eq:uvsum}
\end{equation}
where the summation over $q_j$ has been simply replaced by integration
over $q$, because the phase shifts contribute only a negligible term
proportional to $1/L$.  Substituting Eq.\ (\ref{eq:uvsum}) into Eq.\
(\ref{eq:FUG}) and using Eq.\ (\ref{BCS}), we find
\begin{equation}
   {\delta F_c \over \delta\Delta^*(x)} = 
   {\kappa(\pi/2 -\phi) \over 2\pi\cosh^2(\kappa x)}.
\label{eq:U_c}
\end{equation}
Substituting Eq.\ (\ref{eq:U_c}) and $\partial\Delta^*/\partial\phi$
determined from Eqs.\ (\ref{eq:Delta}) and (\ref{eq:kappa}) into Eq.\
(\ref{eq:dU/dphi}) and integrating over $x$, we get
\begin{equation}
   {dF_c \over d\phi} = -{2\Delta_0 \over \pi}
   \left({\pi\over2} -\phi\right) \sin\phi.
\label{eq:dUdphi}
\end{equation}
Integrating Eq.\ (\ref{eq:dUdphi}) over $\phi$, we recover Eq.\
(\ref{eq:F_c}).

We can also check that the self-consistency condition (\ref{eq:gap})
is satisfied at $\phi=\pi/2$.  Indeed, the last term in the r.h.s.\ of
Eq.\ (\ref{eq:uvsum}) vanishes, and the first term gives
$-\Delta(x)/g$ because of Eq.\ (\ref{BCS}), whereas the contributions
of the localized states
\begin{equation}
   \psi_0^{(\alpha)}(x) = {\sqrt{\kappa} \over 2\cosh\kappa x} 
   \left( \begin{array}{c} 1 \\ -\alpha \end{array} \right)
\label{eq:psi_0}
\end{equation}
mutually cancel for $\alpha=\pm$.

\paragraph{Conclusions.}

We have shown that a 1D superconducting wire with an odd number of
electrons can lower its energy by creating a $\pi$ soliton (kink) of
the order parameter, so that the odd unpaired electron occupies the
midgap bound state localized at the soliton.  In order to exhibit this
effect, the wire does not have to be completely uniform.  It is
sufficient that it has a 1D portion with the length $L$ longer than
the coherence length $\xi$ and the atomic-size cross-section with the
number of conducting channels $N\sim1$ and the transmission
coefficient $\tau\approx1$.  Such atomic Josephson contacts have been
already realized experimentally \cite{Urbina}.  Another possibility is
to use a carbon nanotube \cite{nanotube}.

If the wire is closed in a ring, the phase difference on the two sides
of the soliton would generate a supercurrent in the ring, which could
be detected by a sensitive SQUID.  The total number of electrons in
the ring can be controlled by an external gate in the Coulomb blockade
regime.  \emph{We predict that the supercurrent should appear in the
ring when the number of electrons is odd and disappear when the number
is even.}  The current can flow either clockwise or counterclockwise.
These two degenerate states can be viewed as a qubit and utilized for
quantum computing \cite{qubit}.

Spontaneous current in a ring with phase circulation $\pi$ was
recently predicted for unconventional superconductors using exotic
theory of fractional spin-charge separation \cite{Senthil} and was
disproved experimentally \cite{Moler}.  That theory did not consider
the difference between even and odd number of electrons or the setup
with only one or few conducting channels.  These mesoscopic aspects
play crucial role in our theory, which is formulated completely within
the conventional BCS framework.

\paragraph{Acknowledgments.}

We are grateful to S.\ A.\ Brazovskii, who proposed the idea of this
work, and to D.\ V.\ Averin, F.\ C.\ Wellstood, M.\ S.\ Fuhrer, Y.\
Imry, and R.\ A.\ Webb for useful suggestions and discussion.  The
work was supported by NSF Grant DMR-9815094.

\vspace{-1\baselineskip}


\begin{thebibliography}{00}
\bibitem{current} In the presence of the supercurrent (\ref{eq:J}),
  the phase difference $2\phi$ of the kink slightly deviates from
  $\pi$ producing the current $J=\partial F(\phi)/\partial\phi$
  through the kink \cite{Furusaki,Feigel'man}, which is necessary to
  maintain current continuity.  However, this small deviation does not
  affect the soliton energy $E_s$ significantly if $L\gg\xi$.
\end{thebibliography}
\end{document}